\documentclass[twocolumn,showpacs,preprintnumbers,amsmath,amssymb]{revtex4}

\usepackage{graphicx}
\usepackage{dcolumn}
\usepackage{bm}
\usepackage{soul, xcolor}

\DeclareMathOperator{\D}{d\!}
\DeclareMathOperator{\E}{e} 
\DeclareMathOperator{\I}{i}

\begin{document}

\newtheorem{theorem}{Theorem}
\newtheorem{definition}{Definition}
\newtheorem{lemma}{Lemma}
\newtheorem{proposition}{Proposition}
\newtheorem{remark}{Remark}
\newtheorem{corollary}{Corollary}
\newtheorem{example}{Example}


\title{{The generalized Cattaneo (telegrapher's)  equation and \\ corresponding random walks}}

\author{K. G\'{o}rska}
\email{katarzyna.gorska@ifj.edu.pl}
\author{A. Horzela}
\email{andrzej.horzela@ifj.edu.pl}
\affiliation{Institute of Nuclear Physics, Polish Academy of Science, \\ ul. Radzikowskiego 152, PL-31342 Krak\'{o}w, Poland}

\author{E. K. Lenzi}
\email{eklenzi@uepg.br}
\affiliation{Departamento de Fisica, Universidade Estadual de Ponta Grossa, Av. Carlos Cavalcanti 4748, Ponta Grossa 84030-900, PR, Brazil}

\author{G. Pagnini}
\email{gpagnini@bcamath.org}
\affiliation{BCAM-Basque Centre for Applied Mathematics, \\ 48009 Bilbao, Basque Country Spain and \\ Ikerbasque - Basque Foundation for Science, \\ 48013 Bilbao, Basque Country, Spain}

\author{T. Sandev}
\email{trifce.sandev@manu.edu.mk}
\affiliation{Research Center for Computer Science and Information Technologies, \\ Macedonian Academy of Sciences and Arts, \\ Bul. Krste Misirkov 2, 1000 Skopje, Macedonia}
\affiliation{Institute of Physics \& Astronomy, University of Potsdam, \\ D-14776 Potsdam-Golm, Germany}
\affiliation{Institute of Physics, Faculty of Natural Sciences and Mathematics, \\ Ss Cyril and Methodius University, \\ Arhimedova 3, 1000 Skopje, Macedonia}

\date{\today}

\begin{abstract}
The various types of generalized Cattaneo, called  also telegrapher's equation, are studied. We find conditions under which solutions of the equations considered so far can be recognized as probability distributions, \textit{i.e.} are normalizable and non-negative on their domains. Analysis of the relevant mean squared displacements  enables us to classify diffusion processes described by such obtained solutions and to identify them with either ordinary or anomalous super- or subdiffusion. To complete our study we analyse derivations of just considered examples the generalized Cattaneo equations using the continuous time random walk and the persistent random walk approaches.
\end{abstract}

\maketitle

\section{Introduction}

The standard diffusion (heat) equation describes the collective motion of particles resulting from the random behavior of particles moving in a complex medium. It is a parabolic partial differential equation which unavoidably leads to the infinite propagation velocity and the unphysical conclusion that the action at a distance is possible. The well-known approach proposed to overcome the problem is to consider the hyperbolic differential equation called the telegrapher's or Cattaneo equation \cite{JBKeller04, cattaneo} which standard form is given by 
\begin{equation}\label{classical cattaneo eq}
    T\frac{\partial^{2}}{\partial t^{2}}u(x,t)+\frac{\partial}{\partial t}u(x,t) = A \frac{\partial^{2}}{\partial x^{2}}u(x,t),
\end{equation}
where $T$ is the characteristic time and $A$ is the diffusion coefficient related to the propagation velocity $v=\sqrt{A/T}$. If the initial (localized) condition, \textit{e.g.} given by the Dirac $\delta$-distribution, spreads instantaneously then we deal with the infinite propagation velocity. This is obtained for $T=0$ selected out which means that Eq. \eqref{classical cattaneo eq} becomes the usual diffusion equation 
\begin{eqnarray}\label{classical diffusion eq}
    \frac{\partial}{\partial t}u(x,t) = A \frac{\partial^{2}}{\partial x^{2}}u(x,t).
\end{eqnarray}
Contrary to this, for large $T$, one recovers the wave equation
\begin{eqnarray}\label{classical wave eq}
    \frac{\partial^{2}}{\partial t^{2}}u(x, t)=v^{2}\frac{\partial^{2}}{\partial x^{2}}u(x, t).
\end{eqnarray} 
Therefore, the standard telegrapher's equation is also known as the wave equation with damping. 

To interpret the solution $u(x, t)$ as the probability density function (PDF) it must be non-negative and normalizable function. For such $u(x, t)$ we can calculate the mean square displacement (MSD) $\langle x^{2}(t)\rangle = \int x^{2}u(x,t)\,\D x$ which time dependence allows to recognize the type of diffusion we are dealing with. It is known that the standard Cattaneo equation (\ref{classical wave eq})  in the short time limit yields  the ballistic motion $\langle x^{2}(t) \rangle \propto t^{2}$ and in the long time limit the normal diffusion $\langle x^{2}(t) \rangle \propto t$ \cite{masoliver1,masoliver2,jaume1}. If we want to get anomalous diffusive behavior $\langle x^{2}(t) \rangle \neq t^{\alpha}$, $\alpha > 0$ we should go beyond the standard scheme. A prospective way to proceed is to introduce the generalized Cattaneo equation which inherently involves non-locality in time
\begin{multline}\label{cattaneo like eq}
    \tau \int_{0}^{t}\eta(t - \xi)\frac{\partial^{2}}{\partial \xi^{2}}u_{\eta, \gamma}(x, \xi)\D\xi \\ 
    + \int_{0}^{t}\gamma(t - \xi)\frac{\partial}{\partial\xi}u_{\eta, \gamma}(x, \xi)\D\xi
    = B \frac{\partial^{2}}{\partial x^{2}}u_{\eta, \gamma}(x, t),
\end{multline}
where $\gamma(t)$ and $\eta(t)$ denote memory kernels, while parameters $\tau$ and $B$ are characteristic memory time and generalized diffusion coefficient, respectively. 

It is widely known that equations related to diffusion processes can be derived using various models of the random walk approach. In constructions leading to the Cattaneo equation and its generalizations dominant role is played by the persistent random walk (PRW) and its continuous time extension CTPRW.  Its simplest application provides us with an example of two-state random walk which distinguishes internal states of the random walker - each time the walker is leaving a node $(x,t)$ it moves either left (``plus state'' ) or right (``minus state'') with probability which depends on keeping on or reversing direction of the velocity characterizing the step previously done. Thus, CTPRW based approach opens possibility to include properties of the process which we are interested in - finite propagation velocity and information on the process history. The CTPRW model has been used to derive the standard telegrapher's equation \cite{masoliver1, kac} and the generalized Cattaneo equation with the Caputo derivatives ${_{C}D}^{2\mu}_{t}$ and ${_{C}D}^{\mu}_{t}$ for $0 < \mu < 1$ \cite{masoliver2}. 

In this work we consider examples of the generalized Cattaneo equations which belong to the type of \eqref{cattaneo like eq}. We shall find conditions and/or constraints under which their solutions are normalizable and non-negative. Thereafter, for those solutions we construct the probabilistic interpretation which comes from the CTPRW. The paper is organized as follows. In Sec.~\ref{sec2} we present the basic solution of the standard telegrapher's equation. We also derive fundamental solution of the generalized Cattaneo equation. In Sec.~\ref{sec3} four examples of the generalized Cattaneo equation are presented.  They are distinguished in the Laplace space. In the first two of them the Laplace transformed kernel $\eta(t)$ is equal to the squared Laplace transformed kernel $\gamma(t)$, \textit{i.e.,} $\hat{\eta}(z) = \hat{\gamma}^{2}(z)$. In the next two the Laplace transformed kernels are the same function different from $1$, namely $\hat{\eta}(z) = \hat{\gamma}(z) \neq 1$. We also analyze the moments of the fundamental solution and show that signature of anomalous transport occurs for different forms of the memory kernel. In Sec.~\ref{sec4}, for both cases studied and using formalism of completely monotone, Bernstein, and completely Bernstein functions, we find conditions under which the corresponding solutions are non-negative and, in the consequence, can be called PDFs. In Sec.~\ref{sec5} we formulate sufficient conditions under which the continuous time random walk (CTRW) model coexists with non-negative solutions of relevant diffusion equations. Analogous problem we consider for the persistent random walk model. For generalized Cattaneo equation with $\hat{\eta}(z) = \hat{\gamma}^{2}(z)$ we get the answer which generalizes results of \cite{masoliver2}. The paper is summarized in Sec.~\ref{sec6}.

\section{The standard and generalized Cattaneo (telegrapher's) equations - a glance on the solutions }\label{sec2}

{The standard Cattaneo equation} has been solved analytically in Refs.~\cite{jaume1, jaume2}. For initial conditions
\begin{align}\label{24/07-1}
    u(x,0)=\delta(x), \qquad \frac{\partial}{\partial t}u(x, t)\Big\vert_{t=0}=0
\end{align}
the Fourier-Laplace transform of its solution reads
\begin{align}\label{cattaneo FL}
    \tilde{\hat{u}}(\kappa, z) = \frac{z^{-1}(z + T z^{2})}{(z + T z^{2}) + A \kappa^{2}}.
\end{align}
The arguments $\kappa\in\mathbb{R}$ and $z\in\mathbb{C}$ denote variables in the Fourier and Laplace spaces, while the symbols tilde ``$\sim$" and hat ``$\wedge$" which decorate the function $u(\kappa, z)$ mean the Fourier and Laplace transforms, respectively. The inverse Fourier-Laplace transform of Eq. \eqref{cattaneo FL} leads to Eq.~(32) of Ref.~\cite{jaume1} which for $x_{0} = 0$ yields 
\begin{align}\label{9/01-1}
\begin{split}
    u(x, t) & = \frac{1}{2}\E^{-t/(2T)}\{\delta(x-vt)+\delta(x+vt)\} \\
    & + \frac{\E^{-t/(2T)}}{8 v T}\left[I_{0}(\xi) + \frac{I_{1}(\xi)}{2 T \xi}\right] \Theta(vt - |x|),
\end{split}
\end{align}
where $\xi = \sqrt{v^{2} t^{2} - x^{2}}/(2 v T)$. We point out that $u(x, t)$ given by \eqref{9/01-1} vanished if $|x| > v t$ while for $|x|\leq v t$ is non-negative. The latter is seen from the fact that the modified Bessel functions of the first kind $I_{\nu}(y)$, $\nu \in\mathbb{R}$, are non-negative for $y\ge 0$ as well as the Heaviside step function $\Theta(y)$ and the Dirac $\delta$-distribution. 

For the generalized Cattaneo equation \eqref{cattaneo like eq} we set up the initial conditions \eqref{24/07-1} and calculate its Fourier-Laplace transform. That gives 
\begin{multline}\label{CTRW eta gamma}
    \tau\,\hat{\eta}(z)\left[z^{2}\tilde{\hat{u}}_{\eta, \gamma}(\kappa, z) - z\right] + \hat{\gamma}(z)\left[z\tilde{\hat{u}}_{\eta, \gamma}(\kappa, z)-1\right] \\ 
    = - B \kappa^{2}\tilde{\hat{u}}_{\eta, \gamma}(\kappa, z).
\end{multline} 
Thereafter, we rewrite Eq. \eqref{CTRW eta gamma} in the form
\begin{equation}\label{cattaneo like eq PDF in Laplace-Fourier space M}
    \tilde{\hat{u}}_{\eta, \gamma}(\kappa, z) = \frac{z^{-1}\hat{M}_{\eta, \gamma}(z)}{\hat{M}_{\eta, \gamma}(z) + B \kappa^{2}},
\end{equation}
where 
\begin{equation}\label{M}
   \hat{M}_{\eta, \gamma}(z) = \tau z^{2}\hat{\eta}(z) + z\hat{\gamma}(z).
\end{equation}
The inverse Fourier transform of (\ref{cattaneo like eq PDF in Laplace-Fourier space M}) yields the solution of generalized Cattaneo equation  in the Laplace space
\begin{multline}\label{PDF L}
    \hat{u}_{\eta, \gamma}(x, z) = \frac{z^{-1}[\hat{M}_{\eta, \gamma}(z)]^{1/2}}{2 \sqrt{B}} \\ \times \exp\left(- [\hat{M}_{\eta, \gamma}(z)]^{1/2} \frac{|x|}{\sqrt{B}}\right)
\end{multline}
which, after taking  the inverse Laplace transform of $\hat{u}_{\eta, \gamma}(x, z)$, would provide us with the exact form of $u_{\eta, \gamma}(x, t)$.  If the inverse Laplace transform is not possible to work explicitly out then the Tauberian theorems \cite{feller} become very helpful to find asymptotics of the solutions looked for. 

We have previously mentioned that the important source of physical information of a diffusion process under study is the time dependence of its MSD. The MSD can be calculated twofold employing either $u_{\eta, \gamma}(x, t)$ or $\tilde{\hat{u}}_{\eta, \gamma}(\kappa, z)$ which correspond to each other by the Fourier-Laplace transforms. Calculations done in the Laplace space, \textit{i.e.} using  $\tilde{\hat{u}}_{\eta, \gamma}(\kappa, z)$, seem to be easier in practice. Doing that we represent the MSD as
\begin{equation}\label{30/10-1}
    \langle x^{2}(t)\rangle = \mathcal{L}^{-1}\left.\left[-\frac{\partial^{2}}{\partial\kappa^{2}}\tilde{\hat{u}}_{\eta, \gamma}(\kappa,z)\right]\right|_{\kappa=0}.
    \end{equation}
For the standard Cattaneo equation it gives \cite{jaume1}
\begin{equation}\label{classical cattaneo eq MSD}
    \langle x^{2}(t) \rangle = 2AT\left(t/T - 1 + e^{-t/T}\right).
\end{equation}
Applying the asymptotics of Eq. \eqref{classical cattaneo eq MSD} and using the first three terms in the short time $t \ll 1$ series expansion for  $\exp(-t/\tau)$, \textit{i.e.} $\exp(-t/T) = 1 - t/T + (t/T)^{2} + \ldots$, we get ballistic motion $\langle x^{2}(t)\rangle \propto t^{2}$. In the opposite case $t \gg 1$, $\exp(-t/T)$ decays very fast, so we can estimate Eq. \eqref{classical cattaneo eq MSD} as $\langle x^{2}(t)\rangle \propto t$ which characterizes the normal diffusion. These two behaviors of $\langle x^{2}(t) \rangle $ we pointed out in the Introduction. 

Generalization of the formula \eqref{30/10-1} allows to calculate arbitrary moments 
\begin{align}\label{moments}
\begin{split}
    & \langle x^{n}(t)\rangle_{\eta, \gamma} = \mathcal{L}^{-1}\left[\I^{n}\frac{\partial^{n}}{\partial\kappa^{n}}\tilde{\hat{u}}_{\eta, \gamma}(\kappa, z)\right]_{\kappa=0} \\
    & = \Big\{\begin{array}{c c} (2 m)! B^{m} \mathcal{L}^{-1}\{z^{-1}[\hat{M}_{\eta, \gamma}(z)]^{-m}\}, \quad n=2m, \\
    0, \quad n=2n+1. \end{array}
\end{split}
\end{align}
Obviously, the normalization of $u_{\eta, \gamma}(x, t)$, given by the $0$th moment $\langle x^{0}(t)\rangle$, equals $1$.

\section{Generalized Cattaneo  equations: Special cases}\label{sec3}

Diversity of the recently studied generalized Cattaneo equations concerns their possible applications not solely in physics but also in other branches of science in which detailed understanding of diffusion-like phenomena play an essential role, to mention \textit{e.g.,} chemistry, biology and their mutual interactions. To illustrate this we quote investigations of general properties concerning  anomalous transport in complex media \cite{compte metzler, AGusti18}, some problems in viscoelasticity \cite{YPovstenko11, YPovstenko19} and mote detailed research like \textit{e.g.} description of subdiffusion in a system in which mobile particle $\tilde{A}$ chemically react with static particle $\tilde{B}$ due to the rule $\tilde{A} + \tilde{B} \to \tilde{B}$ \cite{TKosztolowicz14} and characterization of transport processes of electrolytes in subdiffusive media such as gels and porous media \cite{KDLewandowska08}. 

 \subsection*{The memory dependent diffusion and diffusion-wave equations - current state of the art}\label{sub3.1}   

The standard Cattaneo equation is obtained from Eq. \eqref{cattaneo like eq} for $\eta(t) = \gamma(t) = \delta(t)$ which makes it local in time. In what follows we shall discuss analogues of the diffusion equation~\eqref{classical diffusion eq} as well as the wave equation \eqref{classical wave eq} got from Eq. \eqref{cattaneo like eq} if non-trivial memory functions are put into it explicitly.

Eq. \eqref{cattaneo like eq} leads, either for $\eta(t)=0$ or for $\tau\to 0$, to the time smeared diffusion-like equation introduced in \cite{tateishi, sandev metzler chechkin kantz draft}. The latter reads
\begin{eqnarray}\label{diffusion like eq}
    \int_{0}^{t}\gamma(t - \xi)\frac{\partial}{\partial\xi}u_{0, \gamma}(x, \xi)\D\xi = B\frac{\partial^{2}}{\partial x^{2}}u_{0, \gamma}(x, t).
\end{eqnarray}
We note that Eq. \eqref{diffusion like eq} can be derived from the CTRW model with arbitrary waiting time PDF \cite{sandev metzler chechkin kantz draft} or from the over-damped generalized Langevin equation \cite{tateishi}. It is also obtained analyzing the anomalous diffusive process subordinated to normal diffusion under operational time, where the memory kernel is connected to the cumulative distribution function of waiting times \cite{sokolov}.  Conditions under which its solution $u_{0, \gamma}(x, t)$ has probabilistic interpretation, \textit{i.e.,} is non-negative, are discussed in \cite{sandev metzler chechkin kantz draft} (see also \cite{csf2017}).

In \cite{sandev metzler chechkin kantz draft, joint paper submitted} it is shown that the left hand side of Eq. \eqref{diffusion like eq} can be transformed to the distributed order fractional derivative for 
\begin{equation}\label{24/10-6}
    \gamma(t) \equiv \gamma_{1}(t)=\int_{0}^{1} \tau^{\lambda-1}\, p_{1}(\lambda)\frac{t^{-\lambda}}{\Gamma(1-\lambda)} \D\lambda
\end{equation}
with the weight function $p_{1}(\lambda)$ defined on $[0, 1]$ and normalized \textit{i.e.} $\int_{0}^{1} p_{1}(\lambda) \D\lambda = 1$. For $p_{1}(\lambda) = 1$ the Laplace transform of $\gamma_{1'}(t)$ and the auxiliary function $\hat{M}_{0, \gamma_{1'}}(z)$, defined in Eq.\eqref{M} are equal to
\begin{equation}\label{17/01-1}
\hat{\gamma}_{1'}(z) = \frac{z - 1/\tau}{z {\rm ln}(\tau z)} \quad \text{and} \quad \hat{M}_{0, \gamma_{1'}}(z) = \frac{z - 1/\tau}{{\rm ln}(\tau z)}.
\end{equation}
The MSD derived from Eq. \eqref{moments} is given by Eq.~(23) of Ref.~\cite{chechkin2}. Its asymptotic behavior is presented in Eq.~(27) of Ref.~\cite{chechkin2}: the anomalous superdiffusion $\langle x^{2}(t) \rangle_{0, \gamma_{1'}} \propto 2B \tau ( t/\tau) {\rm ln}(\tau/t)$ for the short time limit and ultraslow diffusion $\langle x^{2}(t) \rangle_{0, \gamma_{1'}}  \propto 2 B \tau {\rm ln}(t/\tau)$ for the long time limit.
 
The distributed order fractional derivative approach can be adopted also for the distributed order diffusion equation \cite{chechkin2, chechkin, gorenflo fcaa2013}, the fractional relaxation distributed order \cite{FMainardi06, FMainardi07}, and the diffusion-wave equation in which we set $\gamma(t)=0$. That yields \cite{draft wave eq}
\begin{equation*}
    \tau \int_{0}^{t}\eta(t - \xi)\frac{\partial^{2}}{\partial \xi^{2}}u_{\eta, 0}(x, \xi) \D\xi = B \frac{\partial^{2}}{\partial x^{2}} u_{\eta, 0}(x, t),
\end{equation*}
which can be transformed to the distributed order diffusion-wave equation, considered in \cite{gorenflo fcaa2013}, for 
\begin{equation}\label{24/10-7}
  \eta(t) \equiv \eta_{1}(t) = \int_{1}^{2} \tau^{\lambda-2}\, p_{2}(\lambda)\frac{t^{1-\lambda}}{\Gamma(2-\lambda)} \D\lambda,
\end{equation}
where the weight function $p_{2}(\lambda)$ is normalized on $[1, 2]$. Note that the Laplace transform of $\eta_{1}(t)$ for $p_{2}(\lambda) = 1$ is equal to $\hat{\gamma}_{1'}(z)$ given by Eq. \eqref{17/01-1}. Hence, $\hat{M}_{\eta_{1'}, 0}(z) = \tau z \hat{M}_{0, \gamma_{1'}}(z)$. The asymptotic behaviour of MSD is $\langle x^{2}(t) \rangle_{\eta_{1'}, 0} \propto B\tau (t/\tau)^{2}/2 + B\tau (t/\tau)^{2} {\rm ln}(\tau/t)$ for the short time and $\langle x^{2}(t) \rangle_{\eta_{1'}, 0} \propto 2 B t [{\rm ln}(t/\tau) - 1]$ for the long time limit. The proof of both formulae is in Appendix \ref{app0}.
 
\subsection*{Generalized Cattaneo (telegrapher's) equation: examples of diffusion-wave-like equations}

We shall consider four examples of the generalized Cattaneo equations with non-vanishing kernels $\eta(t)$ and $\gamma(t)$, thus called diffusion-wave equations. Functional forms of these kernels are routinely used in physics: power-law, truncated power-law, and distributed order type.

\smallskip
\noindent
{\bf (A)} The power law memory kernels $\eta_{2}(t) = {t^{1-2\mu}}/{\Gamma(2-2\mu)}$ and $\gamma_{2}(t) = {t^{-\mu}}/{\Gamma(1-\mu)}$ for $\mu\in(0, 1]$ lead to the fractional Cattaneo equation
\begin{multline*}
    \tau^{\mu}\, {_{C}D_{t}^{2\mu}}u_{\eta, \gamma}(x, t) + {_{C}D_{t}^{\mu}}u_{\eta, \gamma}(x, t) \\ = B \frac{\partial^{2}}{\partial x^{2}} u_{\eta_{2}, \gamma_{2}}(x, t),
\end{multline*}
in which the symbols ${_{C}D_{t}^{2\mu}}$ and ${_{C}D_{t}^{\mu}}$ denote the $2\mu$-th and $\mu$-th order fractional derivatives in the Caputo sense \cite{f14/03-1}. The Laplace transform of the memory kernels $\eta_{2}(t)$ and $\gamma_{2}(t)$ satisfy 
\begin{equation*}
\hat{\eta}_{2}(z) = \hat{\gamma}^{2}_{2}(z)\,\,\, \text{with} \,\,\, \hat{\gamma}_{2}(z) = z^{\mu-1}. 
\end{equation*}
The MSD calculated from Eq. \eqref{moments} becomes
\begin{equation}\label{MSD two fractional memory kernels}
    \langle x^{2}(t)\rangle_{\eta_{2}, \gamma_{2}} = 2 B \tau^{\mu}(t/\tau)^{2\mu}E_{\mu, 1+ 2\mu}[-(t/\tau)^{\mu}],
\end{equation}
where $E_{\alpha,\beta}(z)$ is the two parameter Mittag-Leffler function, see Appendix \ref{app-1}. For the short time limit it takes the form $\langle x^{2}(t)\rangle \propto 2 B \tau^{\mu} (t/\tau)^{2\mu}/\Gamma(1+2\mu)$, and for the long time limit is proportional to $2 B t^{\mu}/\Gamma(1+\mu)$. Hence, we conclude that for $0 < \mu \leq 1/2$ the particle spreads out solely in a subdiffusive way while for $1/2 < \mu < 1$ the diffusion changes its character from superdiffusion for the short time  to subdiffusion for the long time. We also pay attention to the fact that for $0 < \mu \leq 1/2$ the maximal order of the derivatives sitting in the equation is less than 1 which preserves its parabolic character and means the infinite propagation velocity.

\smallskip
\noindent
{\bf (B)} The truncated power-law memory kernels $\eta_{3}(t) = \exp(-b t) \eta_{2}(t)$ and $\gamma_{3}(t) = \exp(-b t) \gamma_{2}(t)$ for $b\geq 0$ in the Laplace space read 
\begin{equation*}
\hat{\eta}_{3}(z) = \hat{\gamma}^{2}_{3}(z)\,\,\, \text{with} \,\,\, \hat{\gamma}_{3}(z) = (b + z)^{\mu-1}, \,\, b\geq 0.  
\end{equation*}
The MSD calculated from Eq. \eqref{moments} gives
\begin{align}\label{26/01-3}
\begin{split}
& \langle x^{2}(t) \rangle_{\eta_{3}, \gamma_{3}}
= 2 B \tau \sum_{r\geq 2} (-t^{\mu}/\tau)^{r} E^{(\mu-1)r}_{1, 1 + \mu r}(-b t) \\
&\qquad = 2B \tau \E^{-b t} \sum_{r\geq 2} \frac{ (-t^{\mu}/\tau)^{r}}{\Gamma(1 + \mu r)} {_{1}F_{1}}\!\left({1 + r \atop 1 + \mu r}; b t\!\right),
\end{split}
\end{align}
where $E_{\alpha,\beta}^{\gamma}(z)$ is the three parameter Mittag-Leffler function and ${_{1}F_{1}}\left({a \atop b}; z\right)$ is the hypergeometric function, see Appendices \ref{app-1} and \ref{appA}, respectively. Applying the asymptotics of the three parameter Mittag-Leffler function, see Appendix \ref{app-1}, we find that $\langle x^{2}(t)\rangle_{\eta_{3}, \gamma_{3}} \propto \tau^{1-\mu} \langle x^{2}(t \tau^{1-1/\mu})\rangle_{\eta_{2}, \gamma_{2}}$ for the short time limit and $\langle x^{2}(t)\rangle_{\eta_{3}, \gamma_{3}} \propto \langle x^{2}(t b^{1-\mu}) \rangle$ for the opposite limit. So for $t \ll 1$ it behaves like the MSD in the example {\bf (A)} and for $t \gg 1$ gives the same result as the standard telegrapher's equation.  

\smallskip
\noindent
{\bf (C)} In the next example we take $\eta_{4}(t) = \gamma_{4}(t) = t^{-\mu}/\Gamma(1 - \mu)$, $\mu \in (0, 1]$ which substituting into Eq. \eqref{cattaneo like eq} implies
\begin{multline*}
    \tau{_{C}D_{t}^{1 + \mu}}u_{\eta_{4}, \gamma_{4}}(x, t) + {_{C}D_{t}^{\mu}}u_{\eta_{4}, \gamma_{4}}(x, t) \\ = B \frac{\partial^{2}}{\partial x^{2}} u_{\eta_{4}, \gamma_{4}}(x, t),
\end{multline*}
where ${_{C}D_{t}}^{1 + \mu}$ and ${_{C}D_{t}}^{\mu}$ are fractional derivatives in the Caputo sense. Similarly to {\bf (A)} we get the standard telegrapher's equation for $\mu = 1$. In the Laplace space the kernels $\eta_{4}(t)$ and $\gamma_{4}(t)$ become
\begin{equation*}
\hat{\eta}_{4}(z) = \hat{\gamma}_{4}(z) = z^{\mu - 1}.
\end{equation*}
For such chosen kernels the MSD reads
\begin{equation}\label{MSD two same fractional memory kernels}
    \langle x^{2}(t)\rangle_{\eta_{4}, \gamma_{4}} = 2 B \tau^{\mu} (t/\tau)^{1+\mu} E_{1,2+\mu}(-t/\tau).
\end{equation}
For $0 < \mu < 1$ the particle exhibits diffusion decelerating from  the superdiffusion $\langle x^{2}(t)\rangle \propto 2 B \tau^{\mu} (t/\tau)^{1+\mu}/\Gamma(2+\mu)$ in the short time limit to subdiffusion $\langle x^{2}(t)\rangle \propto 2 B t^{\mu}/\Gamma(1+\mu)$ in the long time limit. 

\smallskip
\noindent
{\bf (D)} The last example comes from the generalized Cattaneo equation \eqref{cattaneo like eq} with $\eta_{1}(t)$ and $\gamma_{1}(t)$.
These kernels are equal if $p_{1}(\lambda) = p_{2}(\lambda + 1) = p(\lambda)$ and their Laplace transform reads 
\begin{align}\label{28/01-1}
\hat{\gamma}_{1}(z) = \hat{\eta}_{1}(z) & = \int_{0}^{1} p(\lambda) (\tau z)^{\lambda - 1} \D\lambda \\
& = \int_{0}^{1} \tilde{p}(\mu) (\tau z)^{-\mu}\D\mu \label{28/01-2}
\end{align}
for $\mu = 1 -\lambda\in (0, 1)$ and $\tilde{p}(\mu) = p(1-\lambda)$ normalized on the range $[0, 1]$. For $\tilde{p}(\mu) = 1$ we have $\hat{\eta}_{1'}(z) = \hat{\gamma}_{1'}(z)$ given by Eq. \eqref{17/01-1}. The asymptotics of MSD $\langle x^{2}(t)\rangle_{\eta_{1}, \gamma_{1}}$ calculated for that case yields
\begin{multline}\label{2/02-12}
\langle x^{2}(t)\rangle_{\eta_{1'}, \gamma_{1'}} \propto 2B\tau [1 + (t/\tau - 1) {\rm ln}(\tau/t)]  - 2B\tau \E^{-t/\tau}\\ 
+ 2 B \tau \E^{-t/\tau} [\gamma - {\rm Ei}(t/\tau)]
\end{multline}
for the short limit and 
\begin{equation}\label{2/02-13}
\langle x^{2}(t)\rangle_{\eta_{1'}, \gamma_{1'}} \propto 2 B \tau {\rm ln}(t/\tau) + 2B \tau \E^{-t/\tau}[\gamma - {\rm Ei}(t/\tau)]
\end{equation}
for the long limit, where ${\rm Ei}(y) = \int_{-y}^{\infty} \exp(-xi)/y \D y$ is the exponential integral. The derivation of Eqs. \eqref{2/02-12} and \eqref{2/02-13} is presented in Appendix \ref{appB}.

\section{Solutions ${u}_{\eta, \gamma}(x, t)$ as the PDFs}\label{sec4}

\subsection*{Completely monotone and Bernstein functions - a brief tutorial}\label{ss41}

The solution $u_{\eta, \gamma}(x, t)$ is the PDF if it is normalizable and non-negative on its domain. Normalizability of $u_{\eta, \gamma}(x, t)$ is easily verified using Eq. \eqref{moments} for $m=0$ but the challenge is to check its non-negativeness. To judge the problem in a possibly general way we shall focus our efforts on developing a method based on the Bernstein theorem which uniquely connects completely monotone functions (abbreviated as CMF) and probability measures through the Laplace integral:  $s\in [0,\infty)\rightarrow G(s) \in \textrm{CMF}$ iff  $G(s) = \int_{0}^{\infty} \exp(-s t) g(t)\!\D t$, $g(t)\ge 0$ for $t\in [0,\infty)$ \cite{book bernstein}. 

We remind that the CMF functions are non-negative functions of a non-negative argument whose all derivatives exist and  alternate, \textit{i.e.}, $(-1)^{n} \D^{\,n} \!G(s)/\!\D s^{n} \geq 0$, $n \in\mathbb{N}$. Note the difference between $G(s)$ and the Laplace transform of $g(t)$, denoted as $\hat{g}(z)$: the first of them is real function of $s > 0$ while the second is complex valued and depends on $z \in \mathbb{C}\setminus\mathbb{R}^{-}$. Knowledge of analytic continuation of $G(s)\rightarrow\hat{g}(z)$ is important  because these are analytical properties of $\hat{g}(z)$ which, according to the Theorem 2.6 of Ref.~\cite{GGripenberg90} quoted as Theorem 1 of Ref.~\cite{ECapelas11}, determine conditions under which $\hat{g}(z)$ is representable as the Laplace transform of a non-negative measure defined on positive semiaxis. The second class of functions which we shall use in this section are Bernstein function (BF) defined as non-negative functions whose derivative is CM:  $h(s)>0$ is BF if $(-1)^{n-1} h^{(n)}(s) \geq 0$, $n = 1, 2, \ldots$ \cite{book bernstein, draft wave eq}. A subclass of BF are complete Bernstein functions (CBF): $c(s)$ is CBF, $s >0$ if $c(s)/s$ is the restricted to the positive semiaxis Laplace transform of a CMF  or, equivalently, the suitably restricted Stieltjes trasform of a positive function \cite{book bernstein, draft wave eq}.  All SFs are completely monotone.

Among the properties of CMFs and BFs which are essential for further considerations we recall that i.) the product of two CMFs is also a CMF and ii.) the composition of CMF and BF function is another CMF \cite{book bernstein}. Consequently,  $\hat{u}_{\eta, \gamma}(x, s)$ given by Eq. \eqref{PDF L} is CMF if $\exp\{-a [\hat{M}_{\eta, \gamma}(s)]^{1/2}\}$, $a > 0$, and $s^{-1} [\hat{M}_{\eta, \gamma}(s)]^{1/2}$ are both CMFs, and $\exp\{-a [\hat{M}_{\eta, \gamma}(s)]^{1/2}\}$ is CMF for $[\hat{M}_{\eta, \gamma}(s)]^{1/2}$ being a BF.  Hence, the non-negativity of $u_{\eta, \gamma}(x, t)$ is ensured by two requirements
\begin{multline}\label{31/01-6}
[\hat{M}_{\eta, \gamma}(z)]^{1/2}\,\,\, \text{is a BF and} \\
s^{-1} [\hat{M}_{\eta, \gamma}(z)]^{1/2} \,\,\, \text{is CMF.}
\end{multline}
The second condition, namely that $s^{-1}[\hat{M}_{\eta, \gamma}(z)]^{1/2}$ is a CMF, immediately follows from the definition of CBF applied to $[\hat{M}_{\eta, \gamma}(z)]^{1/2}$. Thus, two conditions in \eqref{31/01-6} can be replaced by a single one 
\begin{equation}\label{12/03-1}
[\hat{M}_{\eta, \gamma}(z)]^{1/2}\,\,\, \text{is a CBF.}
\end{equation}
To complete needed information on properties of CMF/BF functions we  remind that the composition of BFs is also BF \cite{book bernstein}.

We recall that the non-negativeness of $u_{0, \gamma}(x, t)$ has been shown in \cite{csf2017} with the help of  subordinator approach. However, the subordination does not work for the diffusion-wave equation \cite{draft wave eq} and to analyse properties of $u_{\eta, \gamma}(x, t)$ another tools are needed. Methods of the completely monotone functions, rooted both in classical complex analysis and probability theory,  seem to be the most promising. We shall employ them for $\hat{\eta}(s) = \hat{\gamma}^{2}(s)$ and  $\hat{\eta}(s) = \hat{\gamma}(s)$ using as a starting point either conditions \eqref{31/01-6} or  \eqref{12/03-1}.

\subsection*{The kernels $\hat{\eta}(s) = \hat{\gamma}(s) = 1$}\label{sec4a}

As we mentioned in Sec. \ref{sec2} the solution $u(x, t)$ of the standard telegrapher's equation is non-negative.  It means that $\hat{u}(x, s) = \hat{u}_{1, 1}(x, s)$ is the CMF.  We will repeat this result using the Bernstein theorem applied to $\hat{M}_{1, 1}(s) = T s^{2} + s$. This example confirms our expectations that analysis of functions
$\hat{M}_{1, 1}(s)$ may be helpful also in more complex situations.

To proceed we introduce two functions
\begin{equation}\label{24/10-1}
   F_{1}(T; s) \equiv s^{-1} [\hat{M}_{1, 1}(s)]^{1/2} = T^{1/2}[1 + (T s)^{-1}]^{1/2}
\end{equation}
and 
\begin{equation}\label{24/10-2}
   F_{2}(T; s) \equiv [\hat{M}_{1, 1}(s)]^{1/2} = s F_{1}(T; s)
\end{equation}
which, due to condition \eqref{31/01-6}, should be CMF and BF, respectively. The direct calculation shows that $F_{1}(T; s)$ can be represented by the Laplace integral of the non-negative function $f_{1}(T; t)$ 
\begin{multline*}
    f_{1}(T; t) = \frac{\exp[-t/(2T)]}{2 \sqrt{T}} \Big\{I_{0}\Big(\frac{t}{2T}\Big) + I_{1}\Big(\frac{t}{2T}\Big)\Big\} \\ + \sqrt{T} \delta(t) 
\end{multline*}
which ensures that $F_{1}(T, s)$ is CMF. We point out that the completely monotone character of Eq. \eqref{24/10-1} is also given by \cite[Lemma 1]{KSMiller01}. If $F_{2}(T; s)$ is the BF then $\D F_{2}(T; s)/\D s$ is CMF. Calculating the derivative we find out that it can be obtained by the Laplace integral of
\begin{equation}\label{24/10-3}
    \tilde{f}_{2}(T; t) = \frac{\exp[-t/(2 T)]}{2\sqrt{T}} I_{1}\Big(\frac{t}{2T}\Big) + \sqrt{T} \delta(t).
\end{equation}
Thus, we can say that $\D F_{2}(T; s)/\D s$ is the CMF and, thus, $F_{2}(T; s)$ is the BF.

\subsection*{The kernels $\hat{\eta}(s) = \hat{\gamma}^{2}(s)$}\label{sec4b}

The conditions \eqref{31/01-6} for $\hat{\eta}(s) = \hat{\gamma}^{2}(s)$ are guaranteed for
\begin{equation}\label{26/01-1}
s \hat{\gamma}(s) \,\,\, \text{is a BF and} \,\,\, \hat{\gamma}(s) \,\,\, \text{is a CMF,}
\end{equation}
which can be shown using the results of the previous subsection. For the 
current case the conditions \eqref{31/01-6} read as $s^{-1} [\hat{M}_{\gamma^{2}, \gamma}(s)]^{1/2} = \hat{\gamma}(s) F_{1}[\tau; s \hat{\gamma}(s)]$ and $[\hat{M}_{\gamma^{2}, \gamma}(s)]^{1/2} = F_{2}[\tau; s \hat{\gamma}(s)]$. Assuming that the conditions \eqref{26/01-1} are satisfied we can treat the first and the second requirement in \eqref{31/01-6} as a combination of CMFs or BFs, respectively. If
\begin{equation}\label{12/03-2}
s \hat{\gamma}(s) \,\,\, \text{is CBF}
\end{equation}
we need, as follows from Eq. \eqref{12/03-1}, only one condition to satisfy that $u_{\gamma^{2}, \gamma}(x, t)$ is non-negative function.

For the examples {\bf (A)} we will use only the condition \eqref{12/03-2} whereas for the example {\bf (B)} we will apply the conditions \eqref{26/01-1}. Indeed, for {\bf (A)} we have $s\hat{\gamma}_{2}(s) = s^{\mu}$ which for $0 < \mu \leq 1$ is a CBF and from the definition of CBF appears that $\hat{\gamma}_{2}(s)$ is a CMF. For {\bf (B)} we see that $\hat{\gamma}_{3}(s)$ is a composition of the CMF $\sigma^{\mu-1}$, $\mu\in(0, 1]$, and the BF $\sigma = b + s$. Such composition guarantees that $\hat{\gamma}_{3}(s)$ has completely monotone character. The function $s \hat{\gamma}_{3}(s) = s (b + s)^{\mu - 1}$ is the BF because its first derivative yields 
\begin{equation*}
\frac{\D}{\D s} [s (b + s)^{\mu - 1}] = \mu (b + s)^{\mu - 1} \\ + b (1-\mu) (b + s)^{\mu - 2},
\end{equation*}
which for $0 < \mu \leq 1$ is a CMF. Note that the convex sum of CMFs is another CMF.

\subsection*{The kernels $\hat{\eta}(s) = \hat{\gamma}(s) \neq 1$}\label{sec4c}

Here we consider $\hat{u}_{\eta, \gamma}(x, s)$ for $\hat{\eta}(s) = \hat{\gamma}(s)$. That leads to $\hat{M}_{\gamma, \gamma}(s) = \hat{M}_{1, 1}(s) \hat{\gamma}(s)$. The complete monotonicity 
of $\hat{u}_{\gamma, \gamma}(x, s)$ requires \eqref{31/01-6} to be fulfilled. It is true if
\begin{equation}\label{24/10-4}
    \sqrt{\hat{\gamma}(s)}\,\,\, \text{is CMF and} \,\,\, \sqrt{\tau s^{2}\hat{\gamma}(s) + s \hat{\gamma}(s)}\,\,\, \text{is BF}.
\end{equation}
The function $s^{-1} [\hat{M}_{\gamma, \gamma}(s)]^{1/2}$ is equal to $F_{1}(\tau; s)\, [\hat{\gamma}(s)]^{1/2}$ so its completely monotone character is ensured by the completely monotone character of $[\hat{\gamma}(s)]^{1/2}$. The second conditions, i.e., $[\tau s^{2}\hat{\gamma}(s) + s\hat{\gamma}(s)]^{1/2}$ being BF, is nothing else that the second requirements of Eq. \eqref{31/01-6} saying that $[\hat{M}_{\gamma, \gamma}(s)]^{1/2}$ is BF. 

\begin{figure}[!h]
\includegraphics[scale=0.3]{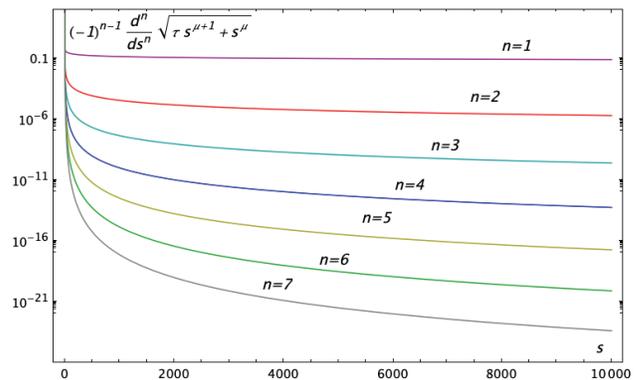}
\caption{\label{fig1} The logarithmic plot of $n$th derivative of $(\tau s^{\mu + 1} + s^\mu)^{1/2}$ multiply by $(-1)^{n-1}$ for $n = 1, 2, \ldots, 7$, $\tau = 1$, and $\mu = 1/2$.}
\end{figure}

As the next example we consider kernels $\eta_{4}(t) = \gamma_{4}(t)$ given in {\bf (C)}. The auxiliary function $\hat{M}_{\gamma_{4}, \gamma_{4}}(s)$ is equal to $F^{2}_{2}(\tau, s) s^{\mu - 1}$. Using the results of the first subsection in Sec.~\ref{sec4a} it is easy to show that $s^{-1} [\hat{M}_{\gamma_{4}, \gamma_{4}}(s)]^{1/2}$ is CMF. Indeed
\begin{align*}
s^{-1} [\hat{M}_{\gamma_{4}, \gamma_{4}}(s)]^{1/2} = s^{-1} F_{2}(\tau, s) s^{(\mu-1)/2} = F_{1}(\tau; s) s^{(\mu - 1)/2}
\end{align*}
which CMF property is ensured by complete monotonicity of $s^{(\mu-1)/2}$ for $\mu \leq 1$, see \cite[Eq. (2)]{book bernstein}. The second condition of Eq. \eqref{24/10-4} for $\hat{\gamma}_{4}(s)$ means that $(\tau s^{\mu + 1} + s^{\mu})^{1/2}$ is BF. Calculating the derivatives we observe that they change the sign according to the rules fulfilled by the BFs. As the confirmation of made calculation we present $(-1)^{n-1} (\D^{\,n}\!\!/\!\D s^{n}) (\tau s^{\mu + 1} + s^{\mu})^{1/2}$ for $\tau = 1$, $\mu = 2/3$, and $n = 1, 2, \ldots, 7$ in Fig. \ref{fig1}.

More complicated is the example {\bf (D)} of Sec.~\ref{sec3} for which the memory kernels are given by Eqs. \eqref{24/10-6} and \eqref{24/10-7}. Due to the condition \eqref{24/10-4} $[\hat{\gamma}_{1}(s)]^{1/2}$ is  CMF and $[\tau s^{2} \hat{\gamma}_{1}(s) + s \hat{\gamma}_{1}(s)]^{1/2}$ is BF. 

The completely monotone character of $[\hat{\gamma}_{1}(s)]^{1/2}$ can be shown as follows: rewrite $[\gamma_{1}(s)]^{1/2}$ as the composition of the CMF $\sigma^{-1/2}$ and $\sigma = [\hat{\gamma}_{1}(s)]^{-1}$ and take into account that the inverse of Eq. \eqref{28/01-2} is BF which is shown in \cite[Subsec. 3.1]{csf2017}. Then applying the fact that the composition of CMF and BF is another CMF completes the proof. 

\begin{figure}[!h]
\includegraphics[scale=0.3]{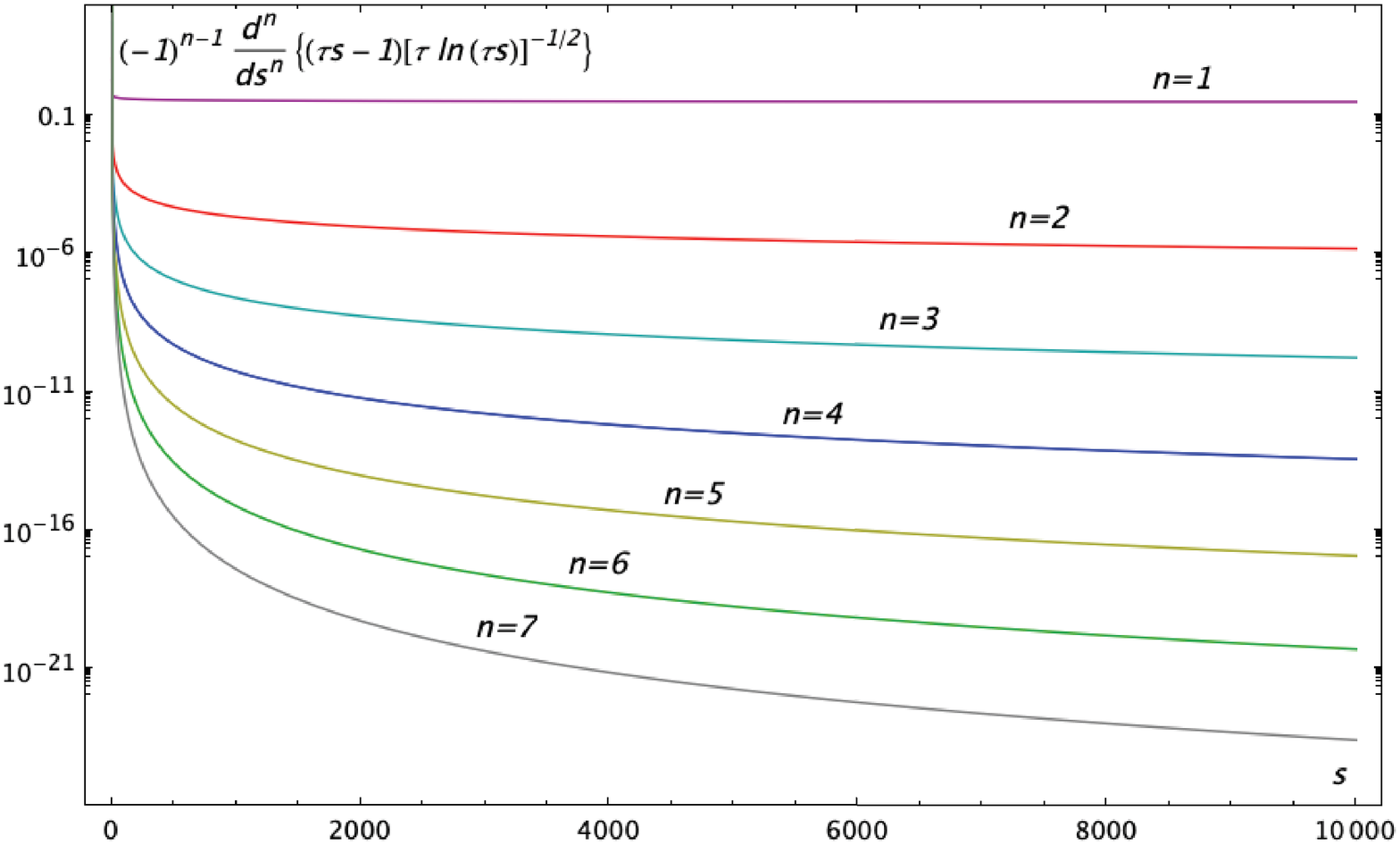}
\caption{\label{fig2} The logarithmic plot of $n$th derivative of $(\tau s - 1)/\sqrt{\tau {\rm ln}(\tau s)}$ multiply by $(-1)^{n-1}$ for $n = 1, 2, \ldots, 7$ and $\tau = 1$.}
\end{figure}

Eq. \eqref{28/01-1} for $\tau = 1$ and $p(\lambda) = 1$ gives $\hat{\gamma}_{1'}(s)$ for $\tau = 1$. Thus, $[s^{2} \hat{\gamma}_{1}(s) + s \hat{\gamma}_{1}(s)]^{1/2} = (s-1)/[{\rm ln}(s)]^{1/2}$ should be BF.  Such statement is confirmed by results of numerical test, exhibited  Fig. \ref{fig2}, which show that $[s^{2} \hat{\gamma}_{1}(s) + s \hat{\gamma}_{1}(s)]^{1/2}$ is non-negative and its odd derivatives and even derivatives are of alternate signs.  

\section{Random walk approaches}\label{sec5}

The CTRW approach generalizes the Brownian motion incorporating two novel features characterizing the random walker motion - the jump length and the waiting time, both obeying  their own probability densities. For this model the probability distribution $P(x,t)$ that the walker is at position $x$ at time $t$, calculated in the Fourier-Laplace space, reads \cite{report_rm}
\begin{equation}\label{7/11-5}
    \tilde{\hat{P}}(\kappa, z) = \frac{1-\hat{\psi}(z)}{z} \frac{1}{1 - \hat{\psi}(z) \tilde{g}(\kappa)}. 
   \end{equation}
Having that we identify $\tilde{\hat{P}}(\kappa, z)$ with the Fourier-Laplace transformed solution of \eqref{CTRW eta gamma} $\tilde{\hat{u}}_{\eta, \gamma}(\kappa, z)$ and assume the standard asymptotic form $\tilde{g}(\kappa) \propto 1 - B \kappa^{2}$ of the Fourier transformed jump density $g(x)$.  These conditions together  with Eq. \eqref{7/11-5} lead to the Laplace transform of the waiting time PDF 
\begin{equation}\label{30/01-2}
\hat{\psi}(z) \equiv \hat{\psi}_{\eta, \gamma}(z) = [1 + \hat{M}_{\eta, \gamma}(z)]^{-1}
\end{equation}
where the subscripts have been added to point out that $\hat{\psi}(z)$ is whenever related to the kernels $\eta(t)$ and $\gamma(t)$ characterizing the particular case of \eqref{CTRW eta gamma} under consideration. The waiting time distribution $\psi(t)$ has to be non-negative which implies that its Laplace transform $\hat{\psi}(z)$ for $z\in {\mathbb R}_{+}$ is to be CMF - sufficient conditions for that may be read out from \eqref{30/01-2} and depend on $1 + \hat{M}_{\eta, \gamma}(z) = 1 + z\hat{\gamma}(z) + \tau z^{2} \hat{\eta}(z)$. To show that $\hat{\psi}_{\eta, \gamma}(z)$ is CMF it is sufficient to prove that $1 + z\hat{\gamma}(z) + \tau z^{2} \hat{\eta}(z)$ is BF. Moreover, because the convex sum of BFs is BF it is enough to check that 
\begin{equation}\label{30/01-3}
z\hat{\gamma}(z) \,\,\, \text{and} \,\,\, z^{2} \hat{\eta}(z) \,\,\, \text{are BFs}. 
\end{equation}

We begin with the kernels $\eta(t)$ and $\gamma(t)$ satisfying in the Laplace space $\hat{\eta}(z) = \hat{\gamma}^{2}(z)$ and notice that Eq. \eqref{26/01-1} implies that $z\hat{\gamma}(z)$ is BF. Thus, the next step is to show that $z^{2} \hat{\gamma}^{2}(z)$ is also BF. The example {\bf (A)} (subscript $2$) says that $z^{2}\hat{\gamma}^{2}_{2}(z) = z^{2\mu}$. It is BF for $0 < \mu \leq 1/2$ so for this range of $\mu$ it can be described using the CTRW. Note that the completely monotone character of $\hat{\psi}_{\eta_{2}, \gamma_{2}}(z)$ for $\mu = 1/2$ can be also explained employing the results of Ref.~\cite{RRNigmatullin16}. For the example {\bf (B)} (superscript 3) $s^{2} \hat{\gamma}_{3}(s)$ can be presented as $s^{2\mu} [1 + 1/(s b^{-1})]^{2(\mu-1)}$ which for $0 < \mu \leq 1/2$ is the product of CMFs and, thus, it is CMF. The fact that $[1 + 1/(s b^{-1})]^{2(\mu-1)}$ is CMF appears from \cite[Lemma 1]{KSMiller01}. In the case of {\bf (C)} (subscript $4$) we have that $z\gamma_{4}(z) = z^{\mu}$ and $z^{2} \hat{\gamma}_{4}(z) = z^{\mu + 1}$. The first one is BF for $0 < \mu \leq 1$, whereas the second is not in the same range of $\mu$. 
Similar situation can be found for case {\bf (D)} where $z \hat{\gamma}_{1}(z)$ is BF (as it is proved in Sec. 2 of Ref.~\cite{csf2017}) but $z^{2} \hat{\gamma}_{1}(z)$ is not BF. In fact, let $h(s) = s \, \hat{\gamma}_{1}(s)$, with $s \in [0,\infty)$, be BF, such that $\D h(s)/\D s = \hat{\gamma}_{1}(s)+s \D\hat{\gamma}_{1}/\D s = G(s)$ is CMF by definition, i.e., $G(s)=\int_{0}^{\infty} \exp(-st) g(t)\D t $ with $g(t) \ge 0$ when $t \in [0,\infty)$, then $\D\,(s^2 \hat{\gamma}_{1})/\D s = h(s) + s G(s)$ is not CMF because $s G(s)$ is not CMF, which implies that $s^2 \hat{\gamma}_{1}(s)$ is not BF. This result follows by observing that $s G(s) = \int_0^\infty \exp(-st) (\D g(t)/\D t) \D t$
is CMF iff $\D g/\D t \ge 0$ but, by setting $s=0$, it holds $\int_0^\infty (\D g/\D t) \D t = 0$ hence $\D g/\D t$  is an oscillating function. 

Just quoted examples demonstrate that coexistence of CTRW and physically admissible solutions of \eqref{CTRW eta gamma} is by no means easy to be achieved. So, when looking for stochastic derivations of  generalized Cattaneo equations we should use another formalism. The method of choice which we are going to advocate and discuss in what follows is the CTPRW model formulated and used to derive the Cattaneo equation in \cite{masoliver0} and next developed in \cite{jaume1, masoliver1, masoliver2, masoliver3} to construct fractional and multidimensional  examples of the generalized Cattaneo equation. We are going to adopt this method in order to study the generalized Cattaneo equation for $\hat{\eta}(z) = \hat{\gamma}^{2}(z)$.  We mentioned in the Introduction that CTPRW is an example of two-state random walk according to which the PDF of finding the random walker at $x$ and $t$ is given by $P(x, t) = P_{+}(x, t) + P_{-}(x, t)$ where $P_{+}(x, t)$ and $P_{-}(x, t)$ denote PDFs that the walker arrives to the location $x,t$ moving either along $+$ or $-$ channel. To construct $P_{\pm}(x, t)$ one introduces the set of joint probability distributions built of the jump $\lambda_{\pm}(x, t)$ and the waiting time $\psi_{\pm}(t)$ PDFs
\begin{equation*}
    h_{\pm}(x, t) = \lambda_{\pm}(x, t)\psi_{\pm}(t),
\end{equation*}
and
\begin{equation*}
    H_{\pm}(x, t) = \lambda_{\pm}(x, t)\Psi_{\pm}(t).
\end{equation*}
In the above $\Psi_{\pm}(t)=\int_{t}^{\infty}\psi_{\pm}(t')\,\D t'$ is the probability that the time which the walker spends at $x$ (called a sojourn) is greater than $t$. Note also that for $\lambda_{\pm}(x, t) = \delta(x\mp v t)$ with $v$ being the speed of the walker we have $\psi_{\pm}(t)=\int_{-\infty}^{\infty}h_{\pm}(x,t)\,\D x$ and $\Psi_{\pm}(t)=\int_{-\infty}^{\infty}H_{\pm}(x,t)\,\D x$. As shown in \cite{masoliver1,masoliver2} the Fourier-Laplace transform of $P(x, t)$ reads
\begin{equation*}
\tilde{\hat{P}}_{\pm}(\kappa, z) = \frac{\tilde{\hat{H}}_{\pm}(\kappa, z)[1 + \tilde{\hat{h}}_{\mp}(\kappa, z)]}{2[1 - \tilde{\hat{h}}_{+}(\kappa, z) \tilde{\hat{h}}_{-}(\kappa, z)]}
\end{equation*}
which, under identification  $\tilde{\hat{P}}(\kappa, z)$
with $\tilde{\hat{u}}_{\eta, \gamma}(\kappa, z)$ and adopting (known or assumed) form of $\tilde{\hat{h}}_{\pm}(\kappa, z)$ and $\tilde{\hat{H}}_{\pm}(\kappa, z)$ may be used to reconstruct the generalized Cattaneo equation \eqref{cattaneo like eq} with relevant functions $\hat{\eta}(z)$ and $\hat{\gamma}(z)$.  
An example is provided by the choice
$\hat{\eta}(z) = \hat{\gamma}^{2}(z)$ and  
\begin{equation}\label{PRW h}
    \tilde{\hat{h}}_{\pm}(\kappa,z) = [1 + 2 \tau z\hat{\gamma}(z)\pm 2\I\!\sqrt{\tau B} \kappa]^{-1},
\end{equation}
\begin{equation}\label{PRW H}
    \tilde{\hat{H}}_{\pm}(\kappa, z) = 2 \tau \hat{\gamma}(z) [1 + 2 \tau z\hat{\gamma}(z)\pm 2\I\!\sqrt{\tau B} \kappa]^{-1}.
\end{equation}

The waiting time PDF $\hat{\psi}_{\pm}(t)$ is related to the Fourier-Laplace transforms of $h_{\pm}(x, t)$ given by Eq. \eqref{PRW h}. Thus, we should calculate the inverse Fourier transform of Eq. \eqref{PRW h} which yields
\begin{equation}\label{7/11-4}
    \hat{h}_{\pm}(x, z) = \frac{\Theta(\pm x)}{2 \sqrt{\tau B}} \exp\left[\mp x \frac{1 + 2 \tau z \hat{\gamma}(z)}{2 \sqrt{\tau B}}\right]
\end{equation}
and using $\hat{\psi}_{\pm}(z) = \int_{-\infty}^{\infty} \hat{h}_{\pm}(x, z) \D x$ we get the functions
\begin{equation}\label{22/01-2}
\hat{\psi}_{+}(z) = \hat{\psi}_{-}(z) = [1 + 2\tau z\hat{\gamma}(z)]^{-1}.
\end{equation}
Treating Eqs. \eqref{7/11-4} and \eqref{22/01-2} as composition of CMF and BF we show that they are CMFs. Employing the fact that convex sum of BFs is another BF, we see that $1 + 2 \tau z \hat{\gamma}(z)$ is a BF, which follow from \eqref{26/01-1} guaranteed that $z \hat{\gamma}(z)$ is a BF. The waiting time PDF computed for the example presented by {\bf (A)} is equal to 
\begin{equation}\label{28/01-3}
\psi_{\pm}(t) = t^{\mu -1 }(2\tau)^{-1} E^{1}_{\mu, \mu}[-t^{\mu}/(2\tau)]
\end{equation}
and they are CMFs for $0 < \mu < 1$, which can be derived from Ref.~\cite{KGorska18}. For the example in {\bf (B)} the waiting time PDF reads
\begin{equation}\label{28/01-4}
\psi_{\pm}(t) = \frac{\E^{-b t}}{t} \sum_{r\geq1} \frac{(-1)^{r + 1}}{\Gamma(r \mu)}\! \left(\frac{t^{\mu}}{2 \tau}\right)^{r}\! {_{1}F_{1}}\left({r \atop r\mu}; bt\right).
\end{equation}
Note that Eqs. \eqref{28/01-3} and \eqref{28/01-4} for $\mu = 1$ lead to the waiting time PDF for the standard telegrapher's equation. Namely, $\psi_{\pm}(t)$ is equal to Eq. (24) of Ref.~\cite{masoliver1} for $\lambda = (2\tau)^{-1}$. Moreover, Eq. \eqref{28/01-4} for $b = 0$ gives Eq. \eqref{28/01-3} because the hypergeometric functions ${_{p}F_{q}}$ at zero argument is equal to 1.

\section{Summary}\label{sec6}

We studied four examples of the generalized Cattaneo equation \eqref{cattaneo like eq} with non-vanishing memory kernels $\eta(t)$ and $\gamma(t)$. Hence, we deal with the differential equation involving the second and first order time derivatives smeared and preserving the shape of space derivative typical for the standard diffusion. In two cases, namely for $\hat{\eta}(s) = \hat{\gamma}^{2}(s)$ and $\hat{\eta}(s) = \hat{\gamma}(s)\neq 1$, we found the conditions under which the solution $u_{\eta, \gamma}(x, t)$ of the generalized Cattaneo equation is the PDF, \textit{i.e.,} is non-negative and normalized function. To show the non-negativity of $u_{\eta, \gamma}(x, t)$ we began with the non-negativity of the solution $u_{1, 1}(x, t) = u(x, t)$ of the standard Cattaneo equation which next provided us the basis of proofs engaging methods of completely monotone and Bernstein functions. Further physical characteristics of the solutions we obtain calculating the moments $\langle x^{n}(t)\rangle = \int x^{n} u_{\eta, \gamma}(x, t) \D x$, $n = 0, 1, 2, \dots$. The normalization of $u_{\eta, \gamma}(x, t)$ is related to the $0$th moment of $u_{\eta, \gamma}(x, t)$ which obviously has to be finite.  Important physical interpretation has the MSD, \textit{i.e.,} the second moment $\langle x^{2}(t) \rangle$ which characterizes the type of diffusion. For $\langle x^{2}(t) \rangle \propto t$ we have the normal diffusion, deviations from this pattern sygnalize anomalous diffusion.  If $\langle x^{2}(t) \rangle$ grows slower that linearly in time $t$ we deal with the sub-diffusion while the opposite case, if occurs,  means the super-diffusion. Physically interesting observation is that the character of anomalous diffusion may change during the time evolution of the system  - analysis of the MSDs time dependence shows that it may behave differently for short and long time limits. Such effects have important consequences if the stochastic processes underlying evolution of the system are taken into account. If the diffusion slows down with increasing time we cannot use the CTRW when are going to derive the generalized Cattaneo equation. Instead, in this case we are forced to use another model of the random walk. The examples {\bf (A)} and {\bf (B)} show that for $0 < \mu \leq 1/2$, where we have the sub-diffusion in both time limits, the waiting time PDF $\psi_{\eta, \gamma}(t)$ of CTRW is non-negative and thus we can adopt the CTRW approach to derive the generalized Cattaneo equation. For $1/2 < \mu \leq 1$ the diffusion slows down from super- to sub-diffusion and $\psi_{\eta, \gamma}(t)$ contains at least one  negative part. This means that we cannot use the CTRW model. An alternative is to take the CTPRW model -  we found a suitable one for $\hat{\eta}(s) = \hat{\gamma}^{2}(s)$. The examples {\bf (C)} and {\bf (D)} \textit{i.e.,} $\hat{\eta}(s)=\hat{\gamma}(s) \ne 1$, also cannot be described by the CTRW and should be analyzed using another approach - CTPRW models are promising challengers also in this case.

\section*{Acknowledgment}
K.G. and A.H. were supported by the NCN, OPUS-12, Program No. UMO-2016/23/B/ST3/01714. K.G. expresses her gratitude to the Macedonian Academy of Sciences and Arts, Skopje, Macedonia, for hospitality during her stay in Skopje and personally to Dr Trifce Sandev for fruitful discussions. T.S. acknowledges support from the Alexander von Humboldt Foundation. G.P. is supported by the Basque Government through the BERC 2018-2021 program and also funded by the Spanish Ministry of Economy and Competitiveness MINECO via the BCAM Severo Ochoa SEV-2017-0718 accreditation.

\appendix

\section{The proof of asymptotics of $\langle x^{2}(t) \rangle_{\eta_{1'}, 0}$}\label{app0}

Let us start with the MSD $\langle x^{2}(t)\rangle_{0,\gamma_{1'}}$ coming from Eq. \eqref{moments}, for which it appears that $\langle x^{2}(t)\rangle_{0,\gamma_{1}} = 2 B \mathcal{L}^{-1}\{[z \hat{M}_{0, \gamma_{1}}(z)]^{-1}\}$ and apply it for the diffusion-wave equation gives
\begin{align}\label{1/02-2}
\begin{split}
\langle x^{2}(t)\rangle_{\eta_{1}, 0} & = 2 B \tau^{-1} \mathcal{L}^{-1}\{[z^{2} \hat{M}_{0, \gamma_{1}}(z)]^{-1}\} \\
& = \tau^{-1} \int_{0}^{t} \langle x^{2}(\xi)\rangle_{0,\gamma_{1}} \D\xi,
\end{split}
\end{align}
where we employ Eq. (1.1.1.7) of \cite{APPrudnikov-v5} from which appears 
\begin{equation}\label{2/02-2}
\mathcal{L}^{-1}[p^{-1} F(p)](y) = \int_{0}^{y} f(\xi) \D\xi,
\end{equation} 
with $f(\xi) = \mathcal{L}^{-1}[F(p)](\xi)$. Employing the asymptotic behaviours of $\langle x^{2}(t)\rangle_{0, \gamma_{1}}$ to Eq. \eqref{1/02-2} we have the asymptotics of MSD of the diffusion-wave equation with distributed order fractional derivative presented in Sec. \ref{sec3}.

\section{The three parameter Mittag-Leffler function}\label{app-1}

The three parameter Mittag-Leffer function is given as the series
\begin{align}\label{30/10-3}
    E^{\gamma}_{\alpha,\beta}(-at^{\alpha})= \sum_{n \geq 0} \frac{(\gamma)_{n} \, (-a t^{\alpha})^{n}}{n! \Gamma(\alpha n+\beta)},
\end{align}
where $\alpha, \beta, \gamma >0$ and $(\gamma)_{n} = \Gamma(\gamma + n)/\Gamma(\gamma) = \gamma (\gamma + 1)\ldots (\gamma + n - 1)$ is the Pochhammer  symbol or rising factorial. It is called the two parameter Mittag-Leffler or the Wiman function for $\gamma = 1$ denoted as $E_{\alpha, \beta}(-at^{\alpha}) = E^{1}_{\alpha, \beta}(-a t^{\alpha})$. For $\gamma = \beta = 1$ we deal with the one parameter (standard) Mittag-Leffler function $E_{\alpha}(- at^{\alpha}) = E_{\alpha, 1}^{1}(-a t^{\alpha})$ whereas it is the exponential function for $\gamma = \beta = \alpha = 1$. At time zero $E^{\gamma}_{\alpha,\beta}(-at^{\alpha})$ is equal to $1/\Gamma(\beta)$. The asymptotic behavior of $E^{\gamma}_{\alpha, \beta}(-a t^{\alpha})$ for the long time limit is $E^{\gamma}_{\alpha,\beta}(-a t^{\alpha})\propto (a^{\gamma} t^{\alpha\gamma})^{-1} /\Gamma(\beta-\alpha\gamma)$  \cite{RGarrappa16, GG}.

The three parameter Mittag-Leffler function multiplied by the power function is known as the Prabhakar kernel $e_{\alpha, \beta}^{\gamma} (-a, t) = t^{\beta - 1} E^{\gamma}_{\alpha,\beta}(-at^{\alpha})$. Its Laplace transform reads 
\begin{equation*}
\mathcal{L}[e_{\alpha, \beta}^{\gamma}(-a, t)] = p^{\alpha\gamma-\beta}(p^{\alpha} + a)^{-\gamma}.
\end{equation*}

\smallskip
\section{The proof of Eq. \eqref{26/01-3}}\label{appA}

Eq. \eqref{moments} for the truncated power-law memory kernels leads to
\begin{multline}\label{30/01-10}
\langle x^{2}(t)\rangle_{\eta_{3}, \gamma_{3}} = 2 B \mathcal{L}^{-1}\{z^{-1}[\tau z^{2} (b+z)^{2\mu - 2}]^{-1} \\
 \times [1 + (\tau z)^{-1} (b + z)^{1 - \mu}]^{-1}\},
\end{multline}
for $b \geq 0$. We apply the series form of $(1 + x)^{-1}$ for $|x| < 1$ to Eq. \eqref{30/01-10} and take that $x = (\tau z)^{-1} (b + z)^{1 - \mu}$. In the considered case the condition $|x| < 1$ means that $|b + z|^{1 - \mu} < \tau |z|$. Thus, we see that 
\begin{equation*}
\langle x^{2}(t)\rangle_{\eta_{3}, \gamma_{3}} = - 2 B \sum_{r\geq 2} (-\tau)^{-(r - 1)} \mathcal{L}^{-1}\!\left[\frac{z^{-(r+1)}}{(b + z)^{r(\mu-1)}} \right],
\end{equation*}
which after using the Laplace transform of the three parameter Mittag-Leffler function written below Eq. \eqref{30/10-3} gives the upper equality in Eq. \eqref{26/01-3}. Employing now Eq. \eqref{30/10-3} and the series for of the hypergeometric function ${_{1}F_{1}}$ due to which 
\begin{equation*}
E_{1, \beta}^{\gamma}(z) = [\Gamma(\beta)]^{-1} {_{1}F_{1}}\left({\gamma \atop \beta}; z\right)
\end{equation*} 
we obtain the lower equality in Eq. \eqref{26/01-3}. 

\section{The proof of Eqs. \eqref{2/02-12} and \eqref{2/02-13}}\label{appB}

Employing Eq. (1.1.1.13) of \cite{APPrudnikov-v5}, namely
\begin{equation}\label{2/02-3}
\mathcal{L}^{-1}[F(p)/(p + a)](y) = \E^{-a y} \int_{0}^{y} f(\xi) \E^{a \xi} \D\xi
\end{equation}
with $f(\xi)$ given below Eq. \eqref{2/02-2}, we can show that $\langle x^{2}(t) \rangle_{\eta_{1}, \gamma_{1}}$ reads
\begin{equation}\label{2/02-1}
\langle x^{2}(t) \rangle_{\eta_{1}, \gamma_{1}} = \tau^{-1} \E^{-t/\tau} \int_{0}^{t} \langle x^{2}(\xi) \rangle_{0, \gamma_{1}} \E^{\xi/\tau} \D\xi.
\end{equation}
For the asymptotic behaviour of $\langle x^{2}(t) \rangle_{0, \gamma_{1}}$ given in Sec. \ref{sec3} and we have Eqs. \eqref{2/02-12} and \eqref{2/02-13}.

\end{document}